\documentclass{iaus}
\usepackage{graphicx}

\title[Correcting the ICFs of bipolar and elliptical PNe] %% give here short title %%
{When Shape Matters: correcting the ICFs to derive the chemical abundances of bipolar 
and elliptical PNe}

\author[Gon\c calves et al.]   %% give here short author list %%
{D. R. Gon\c calves$^{1,2}$, R. Wesson$^{2}$, C. Morisset$^{3, 4}$, 
M. Barlow$^{2}$ \and  B. Ercolano$^{5}$}

 \affiliation{$^{1}$ UFRJ - Observat\'orio do Valongo, Ladeira Pedro Antonio 43, 20080-090 Rio de Janeiro, Brazil.  
 Email: {\tt denise@astro.ufrj.br} \\[\affilskip]
  $^{2}$ Department of Physics and Astronomy, University College London, Gower Street, WC1E
  6BT  London, UK. Email: {\tt rwesson@star.ucl.ac.uk; mjb@star.ucl.ac.uk} \\[\affilskip] 
  $^{3}$ Instituto de Astronom\'\i a, Universidad Nacional Aut\'onoma de M\'exico, Apdo. postal 70-264, 
  04510 Mexico D.F., Mexico Email: {\tt
    chris.morisset@gmail.com}\\[\affilskip]
$^4$Instituto de Astrof\'{\i}sica de Can\'arias, La Laguna, Tenerife, Spain. \\[\affilskip]
  $^{5}$ University Observatory Munich (USM), Scheinerstr 1, D-81679 Muenchen, Germany. \\Email: 
  {\tt ercolano@usm.uni-muenchen.de} \\[\affilskip]
  }

\pubyear{2011}
\volume{283}  %% insert here IAU Symposium No.
\pagerange{ }
\jname{Title of your IAU Symposium}
\editors{A.C. Editor, B.D. Editor \& C.E. Editor, eds.}
\begin{document}

\maketitle

\begin{abstract}
The extraction of chemical abundances of ionised nebulae from a limited spectral range is usually 
hampered by the lack of emission lines corresponding to certain ionic stages. So far, the missing 
emission lines have been accounted for by the ionisation correction factors (ICFs), constructed 
under simplistic assumptions like spherical geometry by using 1-D photoionisation modelling.

In this contribution we discuss the results (Gon\c calves et al. 2011, in prep.) of our ongoing project 
to find a new set of ICFs to determine total abundances of N, O, Ne, Ar, and S, with optical spectra, 
in the case of non-spherical 
PNe. These results are based on a grid of 3-D photoionisation modelling of round, elliptical and 
bipolar shaped PNe, spanning the typical PN luminosities, effective temperatures and densities.

We show that the additional corrections --to the widely used Kingsburgh and Barlow~(1994) ICFs-- 
are always higher for bipolars than for ellipticals. Moreover, these additional corrections are, 
for bipolars, up to: 17\% for oxygen, 33\% for nitrogen, 40\% for neon, 28\% for argon and 50\% for sulphur. 
Finally, on top of the fact that corrections change greatly with shape, they vary also greatly with the central 
star temperature, while the luminosity is a less important parameter.   

%% add here a maximum of 10 keywords, to be taken form the file <Keywords.txt>
\end{abstract}

\firstsection % if your document starts with a section,
              % remove some space above using this command.
\section{A brief history the derivation of nebular abundances}
Accurate ionization correction factors are the key to determining reliable elemental 
abundances for ionized nebulae, for which it is usually the case that only one or two ionization 
stages of a given element can be observed.
 
\cite[Torres-Peimbert \& Peimbert (1979)]{TP79} already proposed and used ICFs to obtain total 
abundances of PNe, a number of years ago. Their ICFs were based on the correlations of the ionic abundances of different elements, like those 
shown in their Fig. 3, for $\log(Ne^{++}/O^{++})$ and $\log(Ar^{++}/O)$ respect to $\log(O^{++}/O)$. 
Other ICF schemes were proposed since then, being the \cite[Kingsburgh \& Barlow (1994)]{KB94}(KB94) the main one.  

The other common approach to derive total abundances is the one in which the empirical (ICF) abundances 
are the input abundances in the photoionisation model fitting of a particular nebula for which a number of other 
input parameters are available (\cite[Stasi\'nska 2002]{ST02}). In this case the empirical abundances are varied until the predicted 
emission-line ratios (and emission-line maps) match the observations (see, for instance, 
\cite[Gon\c calves et al. 2006]{GO06}).                                                 	

\section{Our grid of 3-D models for bipolar and elliptical PNe}

For this study we use {\sc mocassin} (a fully 3-D Monte Carlo photoionisation code): i) which treats the basic ionization and 
recombination (including charge-exchange and dielectronic recombination) physical processes; ii) in which the 
stellar and diffuse radiation fields are treated self-consistently; and iii) is completely independent of 
the assumed nebular geometry (\cite[Ercolano et al. 2003, 2005, 2008)]{ER03,ER05,ER08}. 

\begin{figure}%[b]
% \vspace*{-2.0 cm}
\begin{center}
 \includegraphics[width=5.0in]{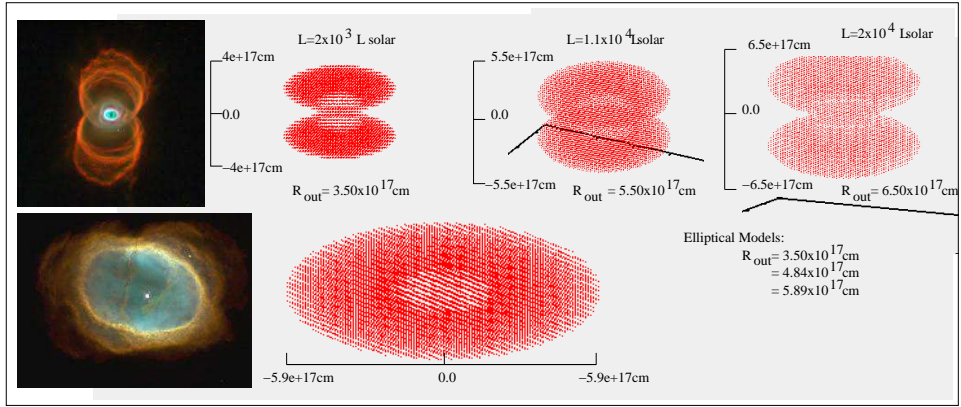} 
% \vspace*{-1.0 cm}
 \caption{The density maps of the 3-D models. The outer radius
of all the 6 B and E geometries adopted are also given. Top: the size
variation with luminosity, for bipolars. Bottom: the largest
elliptical density distribution.}
  % \label{fig1}
\end{center}
\end{figure}

The space parameter to be explored was defined following the previous studies of 
\cite[Alexander \& Balick (1997)]{AB97}, \cite[Perinotto et al. (1998)]{PS98}, 
\cite[Gruenwald \& Viegas (1998)]{GV98} and 
\cite[Morisset \& Stasi\'nska (2008)]{MS08}. We explored blackbody central star temperature ($T_{eff}$) and
 luminosity (L) ranges similar to theirs, with: three different L ($2\times10^3$~L$_{\odot}$, 
$1.1\times10^4$~L$_{\odot}$ and $2\times10^4$~L$_{\odot}$); and varying the $T_{eff}$ from 
$50\times10^3$ to $200\times10^3$~(K) in steps of 25,000~K. The set of chemical abundances of the models 
are two, those of non-type I and type I PNe, as defined by KB94, for He, C, N, O, Ne, Ar and S.  
Bipolar (B) and elliptical (E) bright rim geometries are adopted (see Figure 1), with constant electron densities ($N_e$) of $3\times10^3$~cm$^{-3}$ 
within the shells, and a much higher constant $N_e$ in the torus of the bipolar models 
($N_e$(torus)=6$\times$$N_e$(lobes), \cite[Tafoya et al. 2009]{TA09}).
Finally, the inner and outer radius are varied in order to keep all models radiation bounded. 
As a consequence the higher the luminosity, the larger the nebula. 

\section{Results and highlighting remarks}

\begin{figure}%[b]
% \vspace*{-2.0 cm}
\begin{center}
 \includegraphics[width=5.5in]{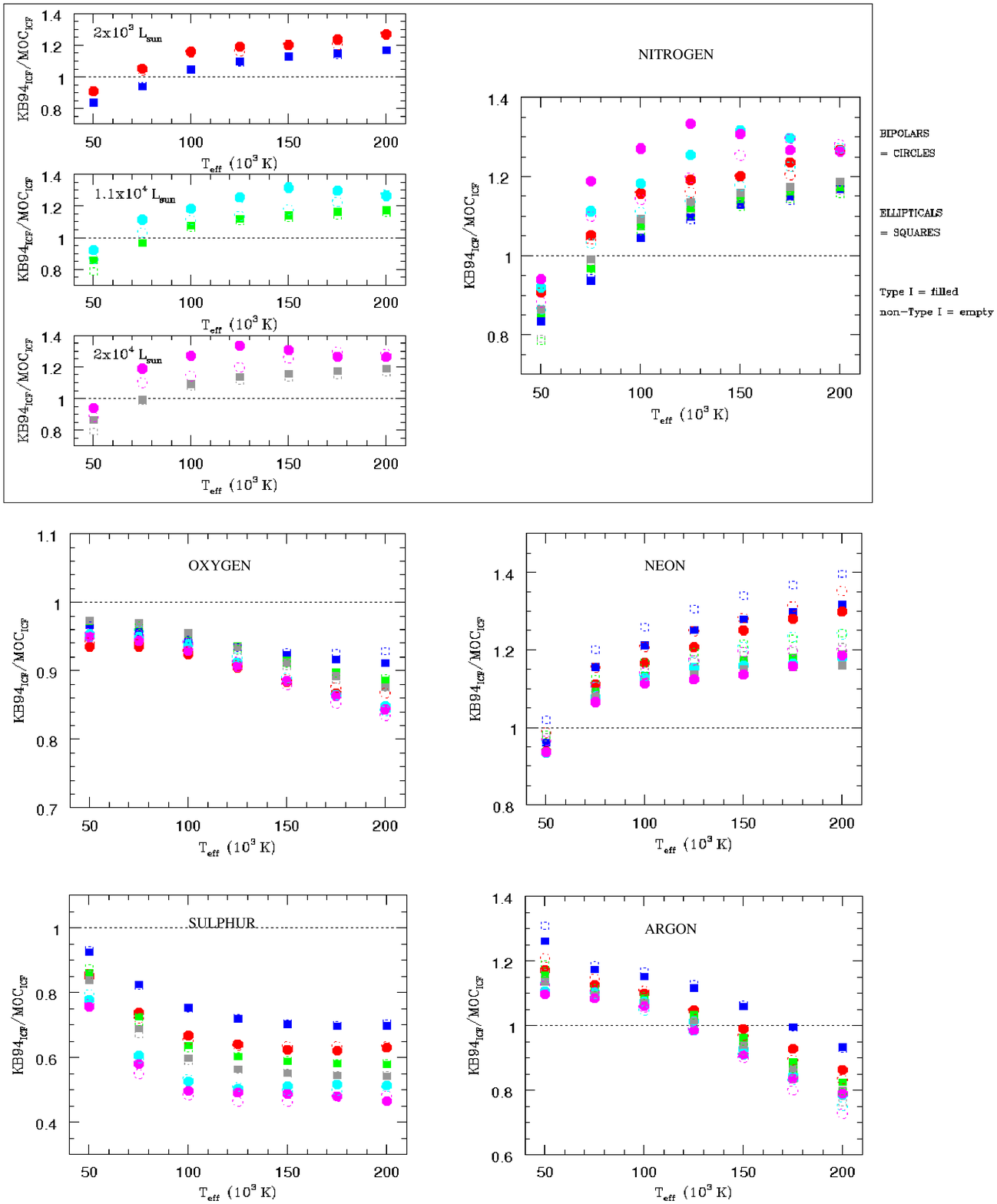} 
% \vspace*{-1.0 cm}
 \caption{The variation of the KB94$_{ICF}$ to MOC$_{ICF}$ ratio for N, O, Ne, S
 and Ar with $T_{eff}$, L, chemical type and morphology. The 3 luminosities are
 show one by one, for nitrogen, in the top-left panel and then all together in
 the top-right one. Red (dark blue), white blue (green) and magenta (grey) represent the 
 $2\times10^3$~L$_{\odot}$, $1.1\times10^4$~L$_{\odot}$ and $2\times10^4$~L$_{\odot}$  
 elliptical (bipolar) models, respectively.
 The same colors apply to the other elements. Squares are for
 bipolar, while circles are for elliptical PNe, being the empty symbols for
 non-type I and the filled ones for type I PNe models. The doted line indicates 
 the total agreement between the two ICFs, therefore points above (below) it indicate over (under) estimation of the true ionization correction factor that would completely account for the total chemical abundances of these elements.}
   %\label{fig1}
\end{center}
\end{figure}

Remembering that no ICFs are needed for He/H (the exception would be the very low-excitation PNe), 
we discuss only the results for O/H, N/H, Ne/H, S/H, and 
Ar/H (hereafter, we use X, instead of X/H, as the total abundance of a given element). 
The reader is referred to the KB94's work to check the set of ICFs we use for the present 
analysis, that is restricted to the equations valid when optical (only) spectra are available. 
Namely, these are equations (A1), (A9), (A28), (A30), 
and (A36) of their Appendix A. 

In our study, first, the line intensities given by {\sc mocassin} 
for the 126 different models (whose parameters are given above) are used to obtain 
the He and the ionic abundances. Second, the KB94's ICF, KB94$_{ICF}$, are calculated based on 
the optical emission-line ratios (ions). Third, 
as {\sc mocassin} models return the ionic fractions, the {\it true} ionization correction factors, 
MOC$_{ICF}$, are obtained. And, finally, the ratio between 
the two ICFs tell us which are the additional corrections we should apply to 
the KB94 scheme in order to derive more robust chemical abundances for bipolar and elliptical PNe. 

Our results are explained in Figure~2. It is straightforward to notice that none of the KB94 ICFs 
can recover the true abundances of bipolar (previously discussed by 
\cite[Gruenwald \& Viegas (1998)]{GV98} in a 3-D fashion, but only for nitrogen) and elliptical PNe. 
For $T_{eff}$ higher than 
$\sim$60$\times10^3$~K, nitrogen and neon are always overestimated by the 
KB94 prescriptions, while oxygen and sulphur are under-estimated within the full range of effective temperatures, and argon goes from over- to under-estimation at about 150$\times10^3$~K. 

We can summarise the main results so far obtained for the discrepancies 
between ICF and true abundances as follows. 

(1) The discrepancies vary with effective temperature as much as with shape, 
and they also change with luminosity and chemical type.

(2) The discrepancies are in general higher for bipolars than for 
   ellipticals.
   
(3) In the worst cases, these discrepancies amount to, for bipolars  (ellipticals): 

\ \ \ \ \ -  up to 33 (19)\% for N (mainly over estimated);

\ \ \ \ \ -  up to 17 (13)\% for O (under estimated);

\ \ \ \ \ -  up to 40 (40)\% for Ne (mainly over estimated);

\ \ \ \ \ -  up to 55 (50)\% for S (under estimated); and 

\ \ \ \ \ -  up to 28 (24)\% for Ar (either under or over estimated).
 
As a matter of fact, we are also studying the ICFs for round (spherically symmetric nebulae), since the approach we adopted here allow us to  
estimated the discrepancies that appear due the variation of L and $T_{eff}$ for simple morphological cases as well. Its clear from our spherically symmetric models that KB94 ICFs should be revisited also in the case of round PNe. 

And, last but not least, the variation of ICF with temperature and luminosity can be more simply 
thought of as a variation with excitation class (although with some scatter). The complete 
discussion, as well as a set of equations for the corrections of the different morphologies, 
will be publish elsewhere soon (Gon\c calves et al. 2011, in prep.).
{}

\end{document}